\documentclass[envcountsame,oribibl]{llncs}

        \makeatletter
        \renewcommand*\l@author[2]{}
        \renewcommand*\l@title[2]{}
        \makeatletter
\usepackage{amsmath,amssymb}
\usepackage{graphicx}
\usepackage{subfigure}
\usepackage{wrapfig}
\usepackage{tikz}
\usetikzlibrary{arrows}
\usetikzlibrary{backgrounds}
\usepackage{amsmath,amssymb}
\usepackage{xcolor}
\usepackage{booktabs}
\usepackage{tikz}
\usepackage{verbatim}
\usepackage{algorithm}
\usepackage{hyperref}
\usepackage{import}
\usepackage{listings}

\newcommand{\K}{{{\cal K}}}
\newcommand{\T}{{{\cal T}}}

\newtheorem{defi}{Definition}
\newtheorem{thm}[defi]{Theorem}

\newtheorem{ex}[defi]{Example}

\newcommand{\ignore}[1]{}

\newcommand{\pK}{\Psi_{\K}}

\usetikzlibrary{matrix,calc,automata,arrows,arrows.meta,backgrounds,positioning,fit}
\tikzstyle{round state}=[circle,draw,solid,fill=white,line width=1pt]
\tikzstyle{square state}=[rectangle,rounded corners,draw,solid,fill=white,line width=1pt]
\tikzstyle{every node}+=[align=center]
\tikzstyle{every picture}+=[remember picture]
\tikzset{accepting/.style={double distance=2pt},%
    every initial by arrow/.style={-{Straight Barb[line width=2pt,length=3mm,width=5mm]}}, %
    initial text=,initial distance=1mm,->,line width=0pt,>=stealth',
    shorten <= 1pt,shorten >=1pt,auto,node distance=2.8cm,
every edge/.style={draw,line width=1pt}}

\newenvironment{enumerate-} 
{\begin{enumerate}
   
   \setlength{\parskip}{-1ex}              
   \setlength{\itemsep}{1.5ex}             
}
{
 \end{enumerate}
}

\begin{document}

\pagestyle{headings}
\title{On the Verification of Parametric Systems}
\titlerunning{On the Verification of Parametric Systems}

 \author{Dennis Peuter, Philipp Marohn and Viorica Sofronie-Stokkermans} 
\authorrunning{Dennis Peuter, Philipp Marohn and Viorica Sofronie-Stokkermans}
\institute{University of Koblenz, Koblenz, Germany\\
$\{$dpeuter,pmarohn,sofronie$\}$@uni-koblenz.de}
\maketitle

\begin{abstract}
We present an approach to the verification of systems for whose description 
some elements -- constants or functions -- are underspecified and can
be regarded as parameters, 
and, in particular, describe a method for automatically generating 
constraints on such parameters under which certain safety conditions are
guaranteed to hold. We present an implementation and illustrate its use
on several examples. 

We dedicate this paper to Werner Damm, whose contributions to the
area of modeling and verification of complex systems were a source of
inspiration for the work described here.
\end{abstract}

\section{Introduction}
Many reasoning problems in mathematics or program
verification can be reduced to
checking satisfiability of ground formulae w.r.t. a theory. 
More interesting however is to 
consider problems -- in mathematics or verification -- in which the
properties of certain function symbols are underspecified
(these symbols are considered to be parametric) and (weakest) additional conditions need
to be derived under which given properties hold.
In this paper we study this type of problems from the perspective of deductive verification: 
We consider parametric reactive and linear hybrid systems -- modeled by transition constraints
or generalizations thereof. 
A classical problem in verification is to check whether a safety property
-- expressed by a suitable formula -- is an invariant, 
or holds for paths of bounded length, {\em for 
given instances of the parameters}, or 
{\em under given constraints on parameters}. 
If the formula expressing the safety property is not 
an inductive invariant, we analyze two possibilities: 
\begin{itemize}
\vspace{-2mm}
\item We present a method that allows us to {\em derive constraints on the parameters}
 which guarantee that the property is an inductive invariant of the system. 
\item We show how this method can be adapted, in certain cases, for {\em strengthening} the formula 
in order to obtain an inductive invariant.
\vspace{-2mm}
\end{itemize}
We present a method for property-directed symbol
elimination in local theory extensions proposed in
\cite{Sofronie-ijcar16,Sofronie-lmcs-2018} and 
illustrate its applicability to solving the tasks above using 
an ongoing implementation in the system \mbox{SEH-PILoT} 
\cite{sehpilot}.

\smallskip
\noindent {\em Related work.} 
Among the approaches to the verification of parametric reactive infinite 
state systems and timed automata we mention 
\cite{Ghilardi, HuneRomijn, Cimatti2}; for parametric hybrid automata
\cite{Henzinger, frehse, farn-wang, Platzer09, Cimatti}. 
However, most papers only consider situations in which the 
parameters are constants. Approaches to 
invariant strengthening or invariant synthesis were proposed in e.g.\ 
\cite{Bradley08,Ghilardi10,Bradley12,Dillig,Kapur06a,Kapur15,Shoham16,Shoham17}.

\smallskip
\noindent In this paper we present a survey of our work in the verification of
parametric systems using hierarchical reasoning and
hierarchical symbol elimination, work which was strongly influenced by
the inspiring collaboration of the last author with Werner Damm in the 
AVACS project and by his papers (e.g.\ \cite{damm-2010, damm-2012}, 
cf.\ also \cite{damm-2018,damm-2020} to mention only a few).
We first used hierarchical reasoning and quantifier elimination for
obtaining constraints on 
the parameters (constants or functions)  in
\cite{sofronie-entcs} where we analyzed possibilities for the verification of
controllers for systems of trains. We then further developed these
ideas in  \cite{Sofronie-ijcar2010}, \cite{sofronie-cade13} and 
\cite{Sofronie-Fundamenta-Informaticae-2020}; in 
\cite{damm-ihlemann-sofronie-2011,damm-ihlemann-sofronie-hscc2011} 
we analyzed possible applications 
to the verification of hybrid systems and, in \cite{peuter-sofronie-cade2019}, 
we showed that similar ideas can also be used e.g.\ for invariant
strengthening (the method for invariant strengthening we proposed extends the results in
\cite{Bradley08,Dillig} to more general theories and is orthogonal to
the approach in \cite{Shoham16,Shoham17}).
After giving an overview of those results, we describe a new,
ongoing, implementation in the system SEH-PILoT, 
and illustrate on some examples how SEH-PILoT can be used for 
the generation of constraints under which a given formula is
guaranteed to be an inductive invariant and for invariant
strengthening.
Another new application is the use of SEH-PILoT for deriving
constraints on parameters under which linear hybrid systems are 
chatter-free.

\smallskip
\noindent 
{\bf Structure of the paper.} 
In Section~\ref{prelim} we give the main definitions needed in the paper
and present existing results on local
theory extensions, 
a method for symbol elimination which turns out to be useful in the verification of
parametric systems, and an implementation.
In Section~\ref{verif} we introduce a class of parametric systems described
by transition constraint systems, we 
 identify situations in which decision procedures 
exist for invariant checking and bounded model checking of such systems, as well 
as methods for obtaining constraints on the parameters which guarantee 
that certain properties are invariants, and methods for invariant
strengthening in transition constraint systems. In Section~\ref{sect-lha} we
study similar problems for some classes of
parametric hybrid automata.
In Section~\ref{conclusions}  we present the conclusions and mention
some plans for
future work.

\smallskip
\noindent This paper is the extended version of
\cite{sofronie-peuter-marohn-2023-vmcai}: it provides 
additional details and examples.

\medskip

\noindent {\bf Table of Contents}

\smallskip

\contentsline {section}{\numberline {1}Introduction}{1}{section.1.1}%
\contentsline {section}{\numberline {2}Local theory extensions}{3}{section.1.2}%
\contentsline {subsection}{\numberline {2.1}Hierarchical reasoning in local theory extensions}{5}{subsection.1.2.1}%
\contentsline {subsection}{\numberline {2.2}Hierarchical symbol elimination}{6}{subsection.1.2.2}%
\contentsline {subsection}{\numberline {2.3}Implementation}{7}{subsection.1.2.3}%
\contentsline {section}{\numberline {3}Verification problems for parametric systems}{8}{section.1.3}%
\contentsline {subsection}{\numberline {3.1}Verification, constraint generation and invariant strengthening}{8}{subsection.1.3.1}%
\contentsline {subsection}{\numberline {3.2}Examples}{11}{subsection.1.3.2}%
\contentsline {section}{\numberline {4}Systems modeled using hybrid automata}{13}{section.1.4}%
\contentsline {subsection}{\numberline {4.1}Verification, constraint generation and invariant strengthening}{15}{subsection.1.4.1}%
\contentsline {subsection}{\numberline {4.2}Chatter-freedom and time-bounded reachability}{15}{subsection.1.4.2}%
\contentsline {subsection}{\numberline {4.3}Examples}{16}{subsection.1.4.3}%
\contentsline {section}{\numberline {5}Conclusions}{21}{section.1.5}%

\section{Local theory extensions}
\label{prelim}

In this section we introduce a class of logical theories used for 
modeling reactive, real time and hybrid systems for 
which we can obtain decidability results. 

\medskip 
\noindent We consider signatures of the form
$\Pi = (\Sigma, {\sf Pred})$ or many-sorted signatures of the form 
$\Pi = (S, \Sigma, {\sf Pred})$, 
where $S$ is a set of sorts, $\Sigma$ is a family of function symbols and ${\sf Pred}$
a family of predicate symbols. 
If $\Pi$ is a signature and $C$ is a set of new constants, we will denote 
by $\Pi^C$ the expansion of $\Pi$ with constants in $C$, i.e.\ 
the signature $\Pi^C = (\Sigma \cup C, {\sf Pred})$.

\noindent We assume known standard definitions from first-order logic.   
In this paper we refer to (finite) conjunctions of
clauses also as ``sets of clauses'', and to (finite) conjunctions of formulae 
as ``sets of formulae''. Thus, if $N_1$ and $N_2$ are finite sets of
formulae then $N_1 \cup N_2$ 
will stand for the conjunction of all formulae in $N_1 \cup N_2$. 
All free variables of a clause (resp. of a set
of clauses) are considered to be universally quantified. 
We denote ``verum'' with $\top$ and ``falsum'' with $\perp$. 

\smallskip
\noindent 
Theories can be defined by specifying a set of axioms, or by specifying a
set of structures (the models of the theory). 
In this paper, (logical) theories are simply sets of sentences.

\smallskip
\noindent If $F, G$ are formulae and ${\mathcal T}$ is a theory we 
write: 
$F \models G$ to express the fact that every model of $F$ is a
  model of $G$; 
$F \models_{\mathcal T} G$ -- also written as ${\mathcal T} \cup F
  \models G$ and sometimes ${\mathcal T} \wedge F \models G$ -- 
to express the fact that every model of $F$ which is also a model of 
$\T$ is a model of $G$.
$F \models \perp$ means that $F$ is
unsatisfiable; $F \models_{\T} \perp$ means that there is no model of
$\T$ in which $F$ is true. 
If there is a model of $\T$ which is also a
model of $F$ we say
that $F$ is satisfiable w.r.t.\ ${\mathcal T}$. 
If $F \models_{\mathcal T} G$ and $G \models_{\mathcal T} F$ we say that
 {\em $F$ and $G$ are equivalent w.r.t.\ ${\mathcal T}$}.

\smallskip
\noindent A theory $\T$ over a signature  
$\Pi$ {\em allows quantifier elimination} if for every formula $\phi$ over  
$\Pi$ there exists a quantifier-free formula $\phi^*$ over  
$\Pi$ which is equivalent to $\phi$ w.r.t.\ $\T$.  
\begin{ex}
Presburger arithmetic with congruence modulo $n$,  
rational linear arithmetic $LI({\mathbb Q})$ and real linear
arithmetic $LI({\mathbb R})$, the theories of real closed fields (real
numbers) and 
of algebraically closed fields, the theory of finite fields, the
theory of absolutely free algebras, and the 
theory of acyclic lists in the signature 
$\{ {\rm car}, {\rm cdr}, {\rm cons} \}$
(\cite{tarski,malcev,chang-keisler,hodges,ghilardi:model-theoretic-methods})
allow quantifier elimination. 
\label{examples-qe}
\end{ex}

\

\noindent {\bf Theory extensions.} 
Let $\Pi_0 {=} (\Sigma_0, {\sf Pred})$ be a signature, and ${\mathcal T}_0$ be a 
``base'' theory with signature $\Pi_0$. We consider 
extensions $\T := {\mathcal T}_0 \cup \K$
of ${\mathcal T}_0$ with new function symbols $\Sigma$
({\em extension functions}) whose properties are axiomatized using 
a set $\K$ of (universally closed) clauses 
in an extended signature $\Pi = (\Sigma_0 \cup \Sigma, {\sf Pred})$,
such that each clause in $\K$ contains function symbols in $\Sigma$. 
Let $\Sigma_P \subseteq \Sigma$ be a set of parameters. 

Let $G$ be a set of ground $\Pi^C$-clauses. We 
want to check whether $G$ is satisfiable w.r.t.\ $\T_0 \cup \K$ or not 
and -- if 
it is satisfiable -- to automatically generate a weakest universal $\Pi_0 \cup 
\Sigma_P$-formula 
$\Gamma$ such that $\T_0 \cup \K \cup \Gamma \cup G$ is 
unsatisfiable. 

\medskip
\noindent In what follows we present situations in which hierarchical reasoning
is complete and weakest constraints on parameters can be computed.

\

\noindent {\bf Local Theory Extensions.} 
Let $\Psi$ be a map which associates with 
every finite set $T$ of ground terms a finite set $\Psi(T)$ of ground terms. 
A theory extension $\T_0
\subseteq \T_0 \cup \K$ is $\Psi$-local if it 
satisfies the condition:

\begin{tabbing}
\= ${\sf (Loc}^\Psi_f)$~ \quad \quad  \= For every finite set $G$ of ground
$\Pi^C$-clauses (for an additional \\
\> \> set $C$ of constants) it holds that $\T_0 \cup {\mathcal K} \cup
G \models \bot$  if and only if \\
\> \> $\T_0
\cup \K[\Psi_\K(G)] \cup G$ is unsatisfiable.
\end{tabbing}

\noindent
where, for every set $G$ of ground $\Pi^C$-clauses,
$\K[\Psi_\K(G)]$ is the set of instances of $\K$ in which the terms
starting with a function symbol in $\Sigma$ are in $\Psi_{\K}(G) = \Psi({\sf
  est}(\K, G))$,
where ${\sf est}(\K, G)$ -- the set of ground \underline{e}xtension \underline{s}ub\underline{t}erms of $\K$ and
$G$ -- is the set of ground terms starting with an
{\em extension function} (i.e.\ a
function in $\Sigma$) occurring in $G$ or $\K$. If $T$ is a set of
ground terms, we use the notation $\Psi_{\K}(T) := \Psi({\sf
  est}(\K, T))$.

In \cite{Sofronie-cade05,ihlemann-jacobs-sofronie-tacas08,Ihlemann-Sofronie-ijcar10} 
we proved that if 
$\Psi_{\K}$ is a term closure operator, i.e.\ the following
conditions hold for all sets of ground terms $T, T'$: 
\begin{enumerate}
\item ${\sf est}(\K, T) \subseteq \pK(T)$,
\item $T \subseteq T' \Rightarrow \pK(T) \subseteq \pK(T')$,
\item $\pK(\pK(T)) \subseteq \pK(T)$,
\item $\pK$ is invariant under constant renaming: for
  any map $h: C \rightarrow C$, $\bar{h}(\pK(T)) = \Psi_{\bar{h}(\K)}(\bar{h}(T))$,
  where $\bar{h}$ is the extension of $h$ to ground terms and formulae.
\end{enumerate}
then $\Psi$-local extensions can be
recognized by showing that certain partial models embed into total
ones.  
Especially well-behaved are theory extensions with the property
$({\sf Comp}^{\Psi}_f)$\footnote{We use the index $f$ in 
$({\sf Comp}_{f})$ in order to emphasize that the property refers to 
completability of partial functions with a finite domain of definition.}  which requires that every partial model of $\T$
whose reduct to $\Pi_0$ is total and the ``set of defined terms'' is
finite and closed under $\Psi$, embeds
into a total model of $\T$ {\em with the same support} (cf. e.g.\ \cite{ihlemann-jacobs-sofronie-tacas08}).
If $\Psi$ is the identity, we denote $({\sf Loc}^{\Psi}_f)$ by $({\sf
  Loc}_f)$ and $({\sf Comp}^{\Psi}_f)$ by $({\sf Comp}_f)$; an
extension satisfying $({\sf  Loc}_f)$ is called {\em local}.
The link between embeddability 
and locality allowed us to identify  many classes of local theory
extensions (cf.\ e.g.\ \cite{Sofronie-cade05,ihlemann-jacobs-sofronie-tacas08,sofronie-ki08}):

\begin{ex}[Extensions with free/monotone functions \cite{Sofronie-cade05,ihlemann-jacobs-sofronie-tacas08}] 
The following types of extensions of a theory $\T_0$ are local: 
\begin{enumerate}
\vspace{-2mm}
\item Any extension of $\T_0$ with uninterpreted function
  symbols ($({\sf Comp}_f)$ holds). 
\item Any extension of a theory ${\cal T}_0$  for which $\leq$ is a partial order 
with functions monotone w.r.t.\ $\leq$ (condition $({\sf
  Comp}_f)$ holds if all models of $\T_0$ are lattices w.r.t.\ $\leq$). 
\vspace{-2mm}
\end{enumerate}
\label{ex-monotone}
\end{ex}
\begin{ex}[Extensions with definitions \cite{JacobsKuncak,ihlemann-jacobs-sofronie-tacas08}]
Consider an extension of a theory $\T_0$ with a new function symbol $f$
defined by axioms of the form: 

\smallskip
$~~~~~~~~{\sf Def}_f := \{ \forall {\overline x} (\phi_i({\overline x}) \rightarrow
    F_i(f({\overline x}), {\overline x})) \mid i =1, \dots, m \}$
    
\smallskip
\noindent (definition by ``case distinction'') where $\phi_i$ and $F_i$, $i = 1, \dots, m$, are formulae
over the signature of $\T_0$  such that the following hold: 
\begin{itemize}
\vspace{-2mm}
\item[(a)] $\phi_i({\overline x}) \wedge \phi_j({\overline x})
  \models_{{\cal T}_0} \perp $  for $i {\neq} j$ and 
\item[(b)] 
${\cal T}_0 \models \forall {\overline x} (\phi_i({\overline x})
  \rightarrow \exists y (F_i(y, {\overline x})))$ for all $i \in \{ 1,
  \dots, m \}$.
\vspace{-2mm}
\end{itemize}
Then the extension is local (and satisfies $({\sf Comp}_f)$). 
Examples: 
\begin{enumerate}
\vspace{-2mm}
\item Any extension with a function $f$ defined by axioms of the form: 

\smallskip
$~~~~~~~~{\sf D}_f := \{ \forall {\overline x} (\phi_i({\overline x}) \rightarrow
 f({\overline x}) = t_i) \mid i = 1, \dots, n \}$   

\smallskip where
 $\phi_i$ are formulae over the signature of $\T_0$ such that (a) holds. 
\item Any extension of $\T_0 \in \{ {\sf LI}({\mathbb Q}), 
{\sf LI}({\mathbb R}) \}$ with functions satisfying 
axioms: 

\smallskip
$~~~~~~~~{\sf Bound}_f := \{ \forall {\overline x} (\phi_i({\overline x}) \rightarrow
 s_i \leq f({\overline x}) \leq t_i) \mid i = 1, \dots, n \}$

\smallskip
\noindent where $\phi_i$ are formulae over the signature of $\T_0$,
$s_i, t_i$ are $\T_0$-terms, condition (a) holds and $\models_{{\cal
    T}_0} \forall
{\overline x} (\phi_i({\overline x}) \rightarrow s_i \leq t_i)$
\cite{ihlemann-jacobs-sofronie-tacas08}.  
\end{enumerate}
\label{examples-local}
\end{ex}

\subsection{Hierarchical reasoning in local theory extensions}

Consider a $\Psi$-local theory extension
${\mathcal T}_0 \subseteq {\mathcal T}_0 \cup {\mathcal K}$.
Condition $({\sf Loc}_f^{\Psi})$ requires that for every finite set $G$ of ground
$\Pi^C$-clauses:
${\mathcal T}_0 \cup {\mathcal K} \cup G \models \perp \text{ if and
  only if  }
{\mathcal T}_0 \cup {\mathcal K}[\Psi_{\mathcal K}(G)] \cup G \models
\bot.$
In all clauses in ${\mathcal K}[\Psi_{\mathcal K}(G)] \cup G$ the function
symbols in $\Sigma$ only have ground terms as arguments, so
${\mathcal K}[\Psi_{\mathcal K}(G)] {\cup} G$ can be flattened
and purified
by introducing, in a bottom-up manner, new
constants $c_t \in C$ for subterms $t {=} f(c_1, \dots, c_n)$ where $f {\in}
\Sigma$ and $c_i$ are constants, together with
definitions $c_t {=} f(c_1, \dots, c_n)$, all included
in a set ${\sf Def}$.
We thus obtain a set of clauses ${\mathcal K}_0 {\cup} G_0 {\cup} {\sf Def}$,
where ${\mathcal K}_0$ and $G_0$ do
not contain $\Sigma$-function symbols and ${\sf Def}$ contains clauses of the form
$c {=} f(c_1, \dots, c_n)$, where $f {\in} \Sigma$, $c, c_1, \dots,
c_n$ are constants.
\begin{thm}[\cite{Sofronie-cade05,ihlemann-jacobs-sofronie-tacas08}]
Let ${\mathcal K}$ be a set of clauses.
Assume that
${\mathcal T}_0 \subseteq {\mathcal T}_1 = {\mathcal T}_0 \cup {\mathcal K}$ is a
$\Psi$-local theory extension.
For any finite set $G$ of ground clauses,
let ${\mathcal K}_0 \cup G_0 \cup {\sf Def}$
be obtained from ${\mathcal K}[\Psi_{\mathcal K}(G)] \cup G$ by flattening and purification,
as explained above.
Then the following are equivalent to ${\mathcal T}_1 \cup G \models \perp$:
\begin{enumerate}
\vspace{-2mm}
\item ${\mathcal T}_0 {\cup} {\mathcal K}[\Psi_{\mathcal K}(G)] {\cup} G \models \perp.$
\vspace{-1mm}
\item ${\mathcal T}_0 \cup {\mathcal K}_0 \cup G_0 \cup {\sf Con}_0 \models \perp,$ where
$\displaystyle{{\footnotesize {\sf Con}_0  {=} \{ \bigwedge_{i = 1}^n c_i
    {=} d_i \rightarrow c {=} d \, {\mid}
\begin{array}{l}
f(c_1, \dots, c_n) {=} c {\in} {\sf Def}\\
f(d_1, \dots, d_n) {=} d {\in} {\sf Def}
\end{array} \}}}.$
\vspace{-2mm}
\end{enumerate}
\label{lemma-rel-transl}
\end{thm}

\vspace{-5mm}
\noindent We can also consider chains of theory extensions:

\smallskip
${\mathcal T}_0 \subseteq  {\mathcal T}_1  =  {\mathcal T}_0 \cup
{\mathcal K}_1  \subseteq  {\mathcal T}_2 = {\mathcal T}_0  \cup
{\mathcal K}_1 \cup {\mathcal K}_2 \subseteq \dots \subseteq {\mathcal
  T}_n = {\mathcal T}_0 \cup {\mathcal K}_1 \cup...\cup {\mathcal K}_n$

\smallskip
\noindent in which each theory is a local extension of the preceding one.
For a chain of $n$ local extensions  a satisfiability check w.r.t.\ the last extension can
be reduced (in $n$ steps) to a satisfiability check w.r.t.\
${\mathcal T}_0$. The only restriction we need to impose in order to
ensure that such a reduction is possible is that at each step the
clauses reduced so far need to be ground -- this is the case 
if each variable in a clause appears at least
once under
an extension function.
This instantiation procedure for chains of local theory
extensions has been implemented in \mbox{H-PILoT} \cite{hpilot}.
\footnote{\mbox{H-PILoT}  allows the user to specify a chain of
  extensions: 
if a function symbol $f$ occurs in ${\mathcal K}_n$ but not in ${\bigcup_{i =
    1}^{n-1}\mathcal K}_i$ 
it is declared as level $n$. }

\subsection{Hierarchical symbol elimination}
\label{sect:symb-elim}
In \cite{Sofronie-lmcs-2018} we proposed a method for
property-directed symbol elimination described in Algorithm~\ref{algorithm-symb-elim}.

\begin{algorithm}[h]
\caption{Symbol elimination in theory extensions
  \cite{Sofronie-ijcar16,Sofronie-lmcs-2018}}
\label{algorithm-symb-elim}
{\small \begin{tabular}{ll}
{\bf Input:} & $\T_0 \subseteq \T_0 \cup \K$ theory extension  with
signature $\Pi = \Pi_0 \cup (\Sigma \cup \Sigma_P)$ \\
                   & ~~where $\Sigma_P$ is a set of parameters and
                   $\K$ is a set of flat clauses. \\
                   & $G$ set of flat ground clauses; $T$ set of flat ground
                   $\Pi^C$-terms s.t.\ ${\sf est}({\cal K}, G) {\subseteq} T$\\
{\bf Output:} & $\forall {\overline y}. \, \Gamma_T({\overline
  y})$ (constraint on parameters; universal $\Pi_0 \cup \Sigma_P$-formula)\\[1ex]
\hline
\end{tabular}
}

{\small
\begin{description}
\vspace{-2mm}
\item[Step 1] Purify $\K[T] \cup G$ as described in Theorem~\ref{lemma-rel-transl} (with set of extension
  symbols $\Sigma_1$).
Let $\K_0 \cup G_0 \cup {\sf Con}_0$ be the set of
  $\Pi_0^C$-clauses obtained this way.

\smallskip
\item[Step 2] Let $G_1 = {\mathcal K}_0 \cup G_0\cup {\sf Con}_0$.
Among the constants in $G_1$, we identify
\begin{enumerate}
\item[(i)] the constants
$c_f$, $f \in \Sigma_P$, where  $c_f$ is a constant
parameter or $c_f$ is
introduced by a definition $c_f = f(c_1, \dots, c_k)$ in the hierarchical
reasoning method, 
\item[(ii)] all constants  ${\overline c_p}$ which are not parameters
  and which 
occur as arguments of functions in $\Sigma_P$ in such definitions.
\end{enumerate}
Replace all the other constants ${\overline  c}$
with existentially quantified variables ${\overline x}$
(i.e.\
replace $G_1({\overline c_p}, {\overline c_f}, {\overline c})$
with $\exists {\overline x}. \, G_1({\overline c_p},
{\overline c_f}, {\overline x})$).

\smallskip
\item[Step 3] Construct
a formula  $\Gamma_1({\overline c_p}, {\overline c_f})$ equivalent to
$\exists {\overline x}. \, G_1({\overline c_p}, {\overline c_f},{\overline
  x})$
w.r.t.\ $\T_0$ using a method for quantifier elimination in
${\mathcal T}_0$ and let $\Gamma_2({\overline c_p, \overline c_f})$ be
$\neg \Gamma_1({\overline c_p}, {\overline c_f})$.

\smallskip
\item[Step 4] Replace
(i) each constant $c_f$ introduced by definition $c_f = f(c_1, \dots,
c_k)$ with the term $f(c_1, \dots,c_k)$ and (ii) ${\overline c_p}$ with universally
quantified variables ${\overline y}$ in $\Gamma_2({\overline c_p}, {\overline
c_f})$. The formula obtained this way is $\forall {\overline y}. \, \Gamma_T({\overline y})$.
\vspace{-2mm}
\end{description}
}
\end{algorithm}

\begin{thm}[\cite{Sofronie-ijcar16,Sofronie-lmcs-2018}]
Let ${\cal T}_0$ be a $\Pi_0$-theory allowing quantifier elimination,
$\Sigma_P$ be a set of parameters
(function and constant symbols) and 
$\Sigma$ a set of function symbols such that $\Sigma \cap (\Sigma_0 \cup \Sigma_P) =
\emptyset$.
Let ${\cal K}$ be a set of flat and linear\footnote{A clause is flat
  if the arguments of extension symbols are variables. A clause is 
  linear if a variable does not occur under different function
  symbols or twice in a term.}
  clauses
in the signature $\Pi_0 {\cup} \Sigma_P {\cup} \Sigma$ in which all
variables occur also below functions in $\Sigma_1 = \Sigma_P \cup
\Sigma$ and $G$ a set of flat ground clauses (i.e.\ the arguments of the
extension functions are constants).
Assume $\T_0 \subseteq \T_0 \cup \K$ satisfies
condition $({\sf Comp}^{\Psi}_f)$ for a suitable closure operator
$\Psi$. 
Let $T = \Psi_{\K}(G)$.
Then Algorithm~\ref{algorithm-symb-elim} yields a universal
$\Pi_0 \cup \Sigma_P$-formula
$\forall {\overline x}. \, \Gamma_T({\overline x})$
s.t.\ ${\cal T}_0 \cup \forall {\overline x}. \,
\Gamma_T({\overline x}) \cup {\cal K} \cup
G \models \perp$, and s.t.\ $\forall {\overline x}. \, \Gamma_T({\overline x})$ is entailed by every
universal formula $\Gamma$ with
  ${\cal T}_0 \cup \Gamma \cup
  {\cal K} \cup G \models \perp$.
\label{thm-alg-1}
\end{thm}
Algorithm~\ref{algorithm-symb-elim} yields a formula $\forall {\overline x}. \,
\Gamma_T({\overline x})$ with
 ${\cal T}_0 \cup \forall {\overline x}. \,
\Gamma_T({\overline x}) \cup {\cal K} \cup
G \models \perp$ also if the extension $\T_0 \subseteq \T_0 \cup \K$ is
not $\Psi$-local
or $T \neq \Psi_{\K}(G)$, but in this case there is no guarantee that $\forall {\overline x}. \,
\Gamma_T({\overline x})$
is the weakest universal formula with this property.

\noindent A similar result holds for chains of local theory
extensions; for details cf.\ \cite{Sofronie-lmcs-2018}.

\subsection{Implementation}
\label{implem}

\noindent {\bf Hierarchical reasoning: \mbox{H-PILoT}.} The method for
hierarchical reasoning in (chains of) local extensions of a base theory described before
was implemented in the system \mbox{H-PILoT} \cite{hpilot}.
\mbox{H-PILoT} 
carries out a hierarchical reduction to the base theory.
Standard SMT provers (e.g.\ CVC4 \cite{cvc} or Z3 \cite{z3-2018}) or 
specialized provers (e.g.\ Redlog \cite{redlog}) are used for testing 
the satisfiability of the formulae obtained after the reduction.
\mbox{H-PILoT} uses eager instantiation, 
so provers like CVC4 or Z3 might in general be faster in proving
unsatisfiability. The advantage of using \mbox{H-PILoT} is that knowing
the instances needed for a complete instantiation allows us to correctly detect
satisfiability (and generate models) in situations in which standard
SMT provers return ``unknown'', and also to use
property-directed symbol elimination
to obtain additional constraints on parameters which
ensure unsatisfiability, as explained in what follows. 

\medskip
\noindent {\bf Symbol elimination: \mbox{SEH-PILoT} (Symbol Elimination with
\mbox{H-PILoT})}.
 For obtaining {\em constraints on parameters}
we use  Algorithm~\ref{algorithm-symb-elim}
\cite{Sofronie-lmcs-2018} which was implemented in
\mbox{SEH-PILoT} (cf. also \cite{sehpilot}) for the case in which the theory can be structured as
a local theory extension or a chain of local theory extensions. 
For the hierarchical reduction, \mbox{SEH-PILoT}  uses H-PILoT. 
The symbol elimination is handled by Redlog. The supported base theories are currently limited
to the theory of real closed fields and the theory of Presburger arithmetic.

\medskip
\noindent {\bf Input.} {\mbox{SEH-PILoT}}  is invoked with an input
file that specifies the tasks and all options. The description of a
task contains: 
\begin{itemize}
\item the description of the problem:  
constraint generation (cf.\ Section~\ref{sect:symb-elim}) or invariant strengthening (cf.\
Section~\ref{invariant-strengthening}); 
\item a list of parameters (or,
alternatively, of symbols to be
eliminated); 
\item optionally a list of conditions on the parameters; 
\item information about the base theory, and about simplification
  options; 
\item the formalization of the actual problem in the syntax of H-PILoT
\footnote{A detailed description of the form of such input files can be
found in~\cite{hpilot}.}.
\end{itemize}

\smallskip
\noindent {\bf Execution.} 
{\mbox{SEH-PILoT}} follows the steps of
Algorithm~\ref{algorithm-symb-elim}.
It uses \mbox{H-PILoT} for the hierarchical reduction and 
writes the result in a file which can be used as input for Redlog.
Optionally, formulae can be simplified using Redlog's interface to the external
QEPCAD-based simplifier SLFQ or with a list of assumptions. The obtained formula
then gets translated from the syntax of Redlog back to the syntax of H-PILoT.
Depending on the chosen mode this is then either the final result of the task
(i.e. a constraint) or the input for the next iteration (i.e.\ invariant
strengthening cf.\ Section~\ref{invariant-strengthening}).

\smallskip
\noindent \textbf{Output:}
The output is a file containing the results of each task. Depending on the
chosen options the file contains in addition a list of the various steps that
have taken place and their results as well as a small statistic showing the
amount of time each step has required and the number of atoms before and after a
simplification was applied.

\section{Verification problems for parametric systems}
\label{verif}
\label{trans}

We identify situations in which 
decision procedures 
for the verification of parametric systems exist and 
in which methods for obtaining constraints on the parameters which guarantee 
that certain properties are invariant can be devised. 

\noindent We specify a reactive 
system $S$ as a tuple 
$(\Pi_S, {\cal T}_S, {\sf Tr}_S)$ where $\Pi_S = (\Sigma_S, {\sf
  Pred}_S)$ is a signature, ${\cal T}_S$ is a 
$\Pi_S$-theory (describing the data types used in the specification
and their properties), and 
${\sf Tr}_S = (V, \Sigma, {\sf Init}, {\sf Update})$ is a transition constraint 
system which specifies:
the variables ($V$) and function symbols ($\Sigma$) 
whose values change over time, where $V \cup \Sigma \subseteq \Sigma_S$; 
a formula ${\sf Init}$ specifying the properties of initial states; 
a formula ${\sf Update}$
with variables in $V {\cup} V'$ and 
function symbols in $\Sigma {\cup} \Sigma'$ (where $V'$ and $\Sigma'$ are 
new copies of $V$ resp.\ $\Sigma$, denoting the variables resp.\ functions 
after the transition) specifying the relationship between the values 
of variables $x$ (functions $f$) before and their values
$x'$ ($f'$) after a transition. 
 
\noindent We consider   
{\em invariant checking} and {\em bounded model checking} problems,
cf.\ \cite{MannaPnueli}:

\smallskip
\noindent {\bf Invariant checking.} 
A formula $\Phi$ is an inductive invariant of a system $S$ with 
theory ${\cal T}_S$ and  
transition constraint system 
${\sf Tr}_S {=} (V, \Sigma, {\sf Init}, {\sf Update})$ if:
\begin{itemize}
\vspace{-1ex}
\item[(1)] $\mathcal{T}_S \wedge {\sf Init} \models \Phi$ and 
\item[(2)] $\mathcal{T}_S \wedge \Phi \wedge {\sf Update} \models \Phi'$,  
where $\Phi'$ results from $\Phi$ by replacing each $x \in V$ by $x'$
and each $f \in \Sigma$ by $f'$.
\end{itemize}
\smallskip
\noindent {\bf Bounded model checking.}
We check whether, for a fixed $k$, states not satisfying a
formula $\Phi$ are 
reachable in at most $k$ steps. Formally, we check whether: 

\noindent 
$\displaystyle{~~~ {\cal T}_S \wedge {\sf Init}_0 \wedge 
\bigwedge_{i = 0}^{j-1} {\sf Update}_i \wedge \neg \Phi_j 
\models \perp \quad \text{ for all } 0 \leq j \leq k}$, 

\noindent 
where ${\sf Update}_i$ is obtained from ${\sf Update}$ by 
replacing every $x {\in} V$  by $x_i$, every  
$f {\in} \Sigma$ by $f_i$, and each 
$x' {\in} V'$, $f' {\in} \Sigma'$ by 
$x_{i+1}, f_{i+1}$; 
${\sf Init}_0$ is ${\sf Init}$ with $x_0$ replacing $x \in V$ and 
$f_0$ replacing $f {\in} \Sigma$; 
$\Phi_i$ is obtained from $\Phi$ similarly.

\subsection{Verification, constraint generation and invariant strengthening}
\label{invariant-strengthening}

We consider transition constraint systems ${\sf Tr}_S = (V, \Sigma, {\sf Init}, {\sf Update})$
in which $\Sigma_S = \Sigma_0 \cup V \cup \Sigma \cup \Sigma_P$, 
the formulae in ${\sf Update}$ contain variables in 
$X$ and functions in $\Sigma$ and possibly parameters in $\Sigma_P$.
We assume that $\Sigma_P \cap \Sigma = \emptyset$.

\medskip 
\noindent We consider universal formulae $\Phi$ which are 
conjunctions of clauses of the form 
$\forall {\overline x} (C({\overline x}, {\overline f}({\overline x}))$, 
where $C$ is a flat clause over $\Sigma_S$.\footnote{We use the following abbreviations: ${\overline x}$ for $x_1, \dots,
x_n$; ${\overline f}({\overline x})$ for $f_1({\overline x}), \dots,
f_n({\overline x})$.} 
Such formulae describe ``global'' properties of the function 
symbols in $\Sigma_S$ at a given moment in time, e.g.\ equality 
of two functions (possibly representing arrays), or monotonicity of a 
function. They can also describe properties of individual elements 
(ground formulae are considered to be in particular universal 
formulae). 
\noindent If the formula $\Phi$ is not an inductive invariant, 
our goals are to: 
\begin{itemize}
\vspace{-1ex}
\item generate constraints on $\Sigma_P$ under which 
  $\Phi$ becomes an inductive invariant;
\item obtain a universally quantified inductive invariant 
$I$ in a specified language 
(if such an inductive invariant exists) such that $I \models_{\T_S}
\Phi$, or a proof that there is no universal inductive 
invariant that entails $\Phi$. 
\end{itemize}
\subsubsection{Verification and constraint generation} ~ \\

\noindent We make the following assumptions: 
Let ${\sf LocSafe}$ be a class of universal formulae over $\Sigma_S$. 
\begin{description}
\vspace{-1ex}
\item[(A1)] There exists a chain of local theory extensions ${\cal
    T}_0 \subseteq \dots \subseteq \T_S \cup {\sf Init}$ such that in
  each extension all variables occur below an extension function.
\item[(A2)] For every $\Phi \in {\sf LocSafe}$ 
there exists a chain of local theory extensions
${\cal T}_0 \subseteq \dots \subseteq \T_S \cup \Phi$ such that in
  each extension all variables occur below an extension function.
\item[(A3)] 
${\sf Update} = \{ {\sf Update}_f \mid f \in F \}$ consists of update
axioms for functions in a set $F$,  where, for every $f
\in F$, ${\sf Update}_f$ has the form 

\smallskip
${\sf Def}_f := \{\forall {\overline x} (\phi^f_i({\overline x}) \rightarrow
C^f_i({\overline x}, f'({\overline x})))\mid i \in I \},$

\smallskip
\noindent such that (i) 
$\phi_i({\overline x}) \wedge \phi_j({\overline x})
  \models_{{\cal T}_S} \perp $  for $i {\neq} j$, 
(ii) ${\cal T}_S \models
  \bigvee_{i = 1}^n \phi_i$,  and (iii) $C^f_i$ are conjunctions of
  literals and 
${\cal T}_S \models \forall {\overline x} (\phi_i({\overline x})
  \rightarrow \exists y (C^f_i({\overline x}, y)))$ for all $i \in I$. 
\footnote{The update axioms 
describe the change of the functions in a set  $F \subseteq \Sigma$, 
depending on a finite set $\{ \phi_i \mid i \in I \}$ of 
mutually exclusive conditions over non-primed symbols. \\
In particular we can consider updates of the form
${\sf D}_{f'}$ 
or ${\sf Bound}_{f'}$ 
 as in Example~\ref{examples-local}.} 
\end{description}
In what follows, for every formula $\phi$ containing 
symbols in $V \cup \Sigma$ we denote by $\phi'$ the
formula obtained from $\phi$ by replacing 
every variable $x \in V$ and 
every function symbol $f \in \Sigma$ with the corresponding symbols $x', f'
\in V' \cup \Sigma'$. 
\begin{thm}[{\small \cite{ihlemann-jacobs-sofronie-tacas08,Sofronie-ijcar2010,Sofronie-Fundamenta-Informaticae-2020}}] 
The following hold under assumptions ${\bf (A1)}-{\bf (A3)}$: 
\begin{itemize}
\item[(1)]  If ground satisfiability w.r.t.\ ${\cal T}_0$ is decidable,
  then for every $\Phi \in {\sf LocSafe}$ (i) the problem of checking
  whether $\Phi$ is an inductive invariant of the system $S$ is decidable; (ii) bounded model
checking $\Phi$ for a fixed bound $k$ is decidable.
\item[(2)] If ${\cal T}_0$ allows quantifier elimination and 
the initial states or the updates contain parameters, the 
symbol elimination method in Algorithm 1 yields constraints on these parameters that guarantee that $\Phi$ is
an inductive invariant. 

If in ${\bf (A1), (A2)}$ we additionally assume that all local
extensions in ${\sf LocSafe}$ satisfy condition $({\sf Comp}_f)$, then
the constraint generated with Algorithm 1 is the weakest among all
universal constraints on
the parameters under which $\Phi$ is an inductive invariant.
\vspace{-2mm}
\end{itemize}
\label{thm-dec-inv-checking-synthesis}
\end{thm}

\subsubsection{Invariant strengthening}~ \\
\label{sect:invgen}

\noindent We now consider the problem of inferring -- in a goal-oriented 
way -- universally
quantified inductive invariants. 
The method we proposed in \cite{peuter-sofronie-cade2019} 
is described in Algorithm~\ref{fig-inv-gen}.

{\small \begin{algorithm}[t]
\caption{Iteratively strengthening a formula to an inductive
  invariant \cite{peuter-sofronie-cade2019}} 
\label{fig-inv-gen}
\begin{tabular}{ll} 
{\bf Input:} & System $S = ((\Sigma_S, {\sf Pred}_S), {\cal T}_S, {\sf
  Tr}_S)$, where ${\sf Tr}_S = (V, \Sigma, {\sf Init}, {\sf Update})$; \\
& $\Sigma_P \subseteq \Sigma_S$; 
$\Phi \in {\sf LocSafe}$ (over $\Sigma_P$) ~\\
{\bf Output:} & 
Inductive invariant $I$ of $S$ that entails $\Phi$ and contains only symbols in $\Sigma_P$\\
&  (if such an invariant exists). \\
\hline 
\end{tabular} 

\begin{tabular}{l} 
1: $I := \Phi$ \\
2: {\bf while} $I$ is not an inductive invariant for $S$ {\bf do:} \\
$~~$ {\bf if} ${\sf Init} \not\models I$ {\bf then} {\bf return} ``no
universal inductive invariant for $S$ over $\Sigma_P$ entails $\Phi$'' \\
$~~$ {\bf if} $I$ is not preserved under ${\sf Update}$ {\bf then} Let
$\Gamma$ be obtained by eliminating \\
$~~$ all primed variables and symbols not in $\Sigma_P$ from $I \wedge {\sf
  Update} \wedge \neg I'$; \\
$~~$ $I := I \wedge \Gamma$\\
3: {\bf return} $I$ is an inductive invariant 
\end{tabular} 
\end{algorithm}}

\medskip
\noindent 
In addition to assumptions {\bf (A1), (A2), (A3)} we now consider the
following assumptions (where $\T_0$ is the base theory in assumptions {\bf (A1)--(A3)}): 
\begin{description}
\vspace{-1mm}
\item[(A4)] Ground satisfiability in ${\cal T}_0$ is decidable; 
  ${\cal T}_0$ allows quantifier elimination. 
\item[(A5)] All candidate invariants $I$ computed in the while loop in 
  Algorithm~\ref{fig-inv-gen} are in
  ${\sf LocSafe}$, and all local extensions in ${\sf LocSafe}$ satisfy
  condition $({\sf Comp}_f)$.
\vspace{-1mm}
\end{description}
Under assumptions $({\bf A1})-({\bf A5})$ 
the algorithm is partially correct:
\begin{thm}[\cite{peuter-sofronie-cade2019}] The following hold: 
\begin{itemize}
\vspace{-2mm}
\item[(1)] If Algorithm~\ref{fig-inv-gen} terminates and returns a formula $I$, then $I$ is an
invariant of the system $S$ containing only function symbols in $\Sigma_P$ that entails $\Phi$. 
\item[(2)] Under assumptions {\bf (A1)}--{\bf (A5)},  
if there exists a universal inductive invariant $J$
containing only function symbols in $\Sigma_P$ that
entails $\Phi$, then 
$J$ entails every candidate invariant $I$ generated in the while
loop of Algorithm~\ref{fig-inv-gen}. 
\item[(3)] Under assumptions {\bf (A1)}--{\bf (A5)},  if
  Algorithm~\ref{fig-inv-gen} terminates because the candidate
  invariant $I$ generated so far is not entailed by ${\sf Init}$ then 
no universal inductive invariant entails $\Phi$.
\vspace{-2mm}
\end{itemize}\label{lemma-entails}
\end{thm}
\begin{thm}[Partial Correctness, \cite{peuter-sofronie-cade2019}]
Under assumptions {\bf (A1)}--{\bf (A5)}, 
if Algorithm~\ref{fig-inv-gen} terminates, then its output
is correct.
\label{inv-gen-correctness}
\end{thm}
In \cite{peuter-sofronie-cade2019} we identified 
situations in which assumption 
({\bf A5}) holds (i.e.\ does not have to be stated explicitly) 
and conditions under which the algorithm terminates.

\subsection{Examples} 

We show how SEH-PILoT can be used for two variants of an example from \cite{peuter-sofronie-cade2019}.

\

\noindent {\bf Example 1:  Invariant checking and constraint generation.} Consider the
following program, using subprogram 
${\sf add1}(a)$, which adds 1 to every element of %
\begin{wrapfigure}{l}{4.2cm}
\centering 
\vspace{-10mm}
{\footnotesize  
\begin{verbatim}
d1 = 1; d2 = 1; i:= 0; 
while (nondet()) { 
   a = add1(a); 
   d1 = a[i]; d2 = a[i+1]; 
   i:= i + 1}
\end{verbatim}
\vspace{-6mm}
}
\label{fig1}
\end{wrapfigure} 
array $a$. 
 We check whether $\Phi := d_2 \geq d_1$ is an 
inductive invariant, and, if not, generate  
additional conditions s.t.\ $\Phi$ is an 
inductive invariant. 

\noindent $\Phi$ holds in the initial states described by  ${\sf Init} := d_1 = 1
\wedge d_2 = 1 \wedge i = 0$; 
it is an inductive invariant of the while 
loop iff the following formula: 

\medskip
\noindent $\begin{array}{@{}l} 
 \forall j (a'[j] = a[j]+1) \wedge d'_1 = a'[i] \wedge d'_2 = a'[i+1]
 \wedge  i' = i + 1 \wedge ~d_1 \leq d_2 ~\wedge~ d'_1 >  d'_2 \\
\end{array}$

\medskip 
\noindent  is unsatisfiable. 
As this formula is satisfiable, $\Phi$ is 
not an inductive invariant. 

\ignore{ 
We illustrate the way we can use \mbox{H-PILoT}  in order to 
make the tests above and \mbox{SEH-PILoT} to obtain the additional condition 
$$\forall i (a[i] \leq a[i+1])$$
under which $\Phi$ is an inductive invariant. 

\

\textbf{Invariant checking.} 
We first check whether the property $\Phi = d_1 \leq d_2$ holds in the
initial state ${\sf Init} := d_1 = 1 \wedge d_2 = 1 \wedge i = 0$ 
using H-PILoT (with external prover Z3):

{\scriptsize 
\begin{lstlisting}[frame=single]
Base_functions := {(+,2), (-,2), (*,2)}
Extension_functions := {(a, 1, 2), (ap, 1, 3)}
Relations := {(<=,2), (<,2), (>=,2), (>,2)}

Clauses := % --INIT--
           d1 = _1;  d2 = _1; i = _0;
Query :=   d1 - d2 > _0; % --not Psi(d1, d2)-- 
\end{lstlisting}
}

\smallskip 
\noindent H-PILoT detects unsatisfiability. This shows that $\Phi$ is true in the initial state. 
Next we check whether $\Phi$ is invariant under updates. We again use H-PILoT (with external prover Z3) for this:

{\scriptsize 
\begin{lstlisting}[frame=single]
Base_functions := {(+,2), (-,2), (*,2)}
Extension_functions := {(a, 1, 2), (ap, 1, 3)}
Relations := {(<=,2), (<,2), (>=,2), (>,2)}

Clauses := % --Psi(d1, d2)--
            d1 <= d2;
            % --Update--
                 (FORALL j). ap(j) = a(j) + _1;
                 d1p = ap(i);  d2p = ap(i + _1); ip = i + _1;
Query :=   d1p - d2p > _0;  % --not Psi(d1p, d2p)--
\end{lstlisting}
}

\smallskip 
\noindent H-PILoT cannot detect unsatisfiability. This shows that the 
property is not an inductive invariant yet. 
}

\

\noindent {\bf Constraint generation.} We use SEH-PILoT to generate constraints on the parameters $\Sigma_P =
\{ a, d_1, d_2 \}$ under which $\Phi$ is an inductive invariant. 

{\scriptsize 
\begin{lstlisting}[frame=single]
tasks:
    example constraint generation:
        mode: GENERATE_CONSTRAINTS 
        options:
            parameter: [a, d1, d2]
            slfq_query: true 
        specification_type: HPILOT 
        specification_theory: REAL_CLOSED_FIELDS 
        specification:
            file: |
                Base_functions := {(+,2), (-,2), (*,2)}
                Extension_functions := {(b, 1, 1), (a, 1, 2), (ap, 1, 3)}
                Relations := {(<=,2), (<,2), (>=,2), (>,2)}
                
                Clauses := % --Phi(d1,d2)-
                    d1 <= d2;
                    % --Update--
                    (FORALL j). ap(j) = a(j) + _1;
                    d1p = ap(i); 
                    d2p = ap(i + _1);
                    ip = i + _1;
                Query :=   d1p - d2p > _0; %--not Psi(d1p,d2p)--

\end{lstlisting}
}

\smallskip 
\noindent 
SEH-PILOT gives the following output:

{\scriptsize \begin{lstlisting}[frame=single]
Metadata:
    Date: '2023-07-24 15:04:48' 
    Number of Tasks: 1 
    Runtime Sum: 0.4339 
example constraint generation:
    Runtime: 0.4339 
    Result: (FORALL i). OR(a(i + _1) - a(i) >= _0, d1 - d2 > _0) 
\end{lstlisting}
}

\smallskip 
\noindent Since {\tt ap} (i.e. $a'$) is defined by a clause 
satisfying the requirements for an extension by definitions, it 
defines a local extension satisfying $({\sf Comp}_f)$, therefore  
$\forall i (a[i] {\leq} a[i{+}1])$ is the weakest condition under which 
$\Phi$ is an inductive invariant.

\

\noindent {\bf Example 2:  Invariant strengthening.} 
Consider the program below, using subprograms
${\sf copy}(a, b)$,  which copies the array $b$ into array $a$, 
and ${\sf add1}(a)$, which adds 1 to every element of array $a$. 
The task is to prove that if $b$ is an array with 
\begin{wrapfigure}{l}{4.2cm}
\centering 
\vspace{-10mm}
{\footnotesize  
\begin{verbatim}
d1 = 1; d2 = 1; i:= 0; 
copy(a, b); 
while (nondet()) { 
   a = add1(a); 
   d1 = a[i]; d2 = a[i+1]; 
   i:= i + 1}
\end{verbatim}
}\vspace{-10mm}

\label{fig2}
\end{wrapfigure} 
its  elements sorted  in  increasing order then  the formula $\Phi := d_2 \geq d_1$ is an
invariant of the program. 
It can be checked that 
$\Phi$ holds in the initial states ${\sf Init} ~:=~ d_1 {=} 1 ~\wedge~ d_2 {=}
1 ~\wedge~ i {=} 0 ~\wedge$ $\forall l (a(l) = b(l)) \wedge \forall l, j (l \leq j
\rightarrow b(l) \leq b(j))$. We can prove that it is not an inductive invariant. 

\vspace{2mm}

\noindent We strengthen $\Phi$ using SEH-PILoT.  The tests using, this
time, the
theory of Presburger arithmetic are included below (the result is the
same also if we use as base theory the theory of real closed fields).

{\scriptsize 
\begin{lstlisting}[frame=single]
sehpilot_options:
    keep_files: true

task_options:
    print_steps: true

tasks:
    example_4.16:
        mode: INVARIANT_STRENGTHENING
        options:
            inv_str_max_iter: 2
            parameter: [a, d1, d2]
        specification_type: PTS
        specification_theory: PRESBURGER_ARITHMETIC
        specification:
            base_functions: "{(+,2), (-,2), (*,2)}"
            extension_functions: "{(b, 1, 1), (a, 1, 2), (ap, 1, 3)}"
            relations: "{(<=,2), (<,2), (>=,2), (>,2)}"
            init: |
                d1 = _1; 
                d2 = _1;
                (FORALL j). a(j) = b(j);
                i = _0;
                (FORALL i,j). i <= j --> b(i) <= b(j);
            update: |
                (FORALL j). ap(j) = a(j) + _1;
                d1p = ap(i); 
                d2p = ap(i + _1);
                ip = i + _1;
            query: |
                d1 <= d2;
            update_vars:
                a : ap
                d1 : d1p
                d2 : d2p
                i : ip
\end{lstlisting}}

\smallskip
\noindent Because of the flag ``{\tt print\_steps: true}'', SEH-PILOT prints information
about all steps; because of the flag ``{\tt inv\_str\_max\_iter: 2}'', it sets
  2 as an upper limit for the number of iterations. 
Then, the list of parameters is provided and the base theory (Presburger
arithmetic) is specified. The {\em transition system} itself is then
specified by: 
\begin{itemize} 
\item describing the signature of the base theory; 
\item describing the chain of theory extensions by specifying the
  extension functions and the level of the extension; 
\item describing the initial states by a set of formulae {\tt init}; 
\item describing the updates using a set of formulae  {\tt update}; 
\item specifying the candidate invariant, i.e.\ the formula $\Phi :=
  d_1 \leq d_2$.
\end{itemize}
In addition, under ``{\tt update\_vars}'', the relationship between
the name of the variables
  and function symbols before and after the update is described: 
The variables and function symbols which are updated are $a$ ({\tt
  a}), $d_1, d_2$ ({\tt d1,
  d2}) and $i$ ({\tt i});  the symbols used for the values after the update are ($a'$ ({\tt
  ap}),
$d'_1, d'_2$ ({\tt d1p, d2p}) and resp.\ $i'$ ({\tt ip})).

\medskip 
\noindent 
SEH-PILOT gives the following output:

\

{\scriptsize \begin{lstlisting}[frame=single]
Metadata:
    Date: '2023-10-26 13:15:50'
    Number of Tasks: 1
    Runtime Sum: 0.7239
example_4.16:
    Result:
        Inductive Invariant: |-
            d1 - d2 <= _0;
            (FORALL i). a(i + _1) - a(i) >= _0;
    Runtime: 0.7239
    Extra:
        1. Iteration:
            (step) current candidate: d1 <= d2;
            (step) negated candidate: d1 - d2 > _0;
            (step) negated and updated candidate: d1p - d2p > _0;
            (step) created subtask:
                name: example_4.16_ST_strengthening_1_
                mode: Mode.SYMBOL_ELIMINATION
            (step) verification condition init: true
            (step) verification condition: false
            (step) new candidate: |-
                d1 - d2 <= _0;
                (FORALL i). a(i + _1) - a(i) >= _0;
        2. Iteration:
            (step) current candidate: |-
                d1 - d2 <= _0;
                (FORALL i). a(i + _1) - a(i) >= _0;
            (step) negated candidate: 
                                  OR(a(sk_i + _1) - a(sk_i) < _0, 
                                       d1 - d2 > _0);
            (step) negated and updated candidate: 
                                  OR(ap(sk_i + _1) - ap(sk_i) < _0, 
                                       d1p - d2p > _0);
            (step) created subtask:
                name: example_4.16_ST_VC_update_2
                mode: Mode.GENERATE_CONSTRAINTS
            (step) verification condition init: true
            (step) verification condition: true
\end{lstlisting}
}

\noindent We obtain: 

\smallskip
$~~~~~~\Gamma = \forall i (a(i) \leq a(i+1) \vee d_1 > d_2)$

\smallskip
\noindent Combining the constraint computed by SEH-PILoT with $\Phi$
we obtain the following candidate invariant:

\smallskip
$~~~~~~\Phi_1 = d_1 \leq d_2 \land \forall i ~ a(i+1) \geq a(i)$

\smallskip
\noindent To check whether $\Phi_1$ is an inductive
invariant, we first check whether it holds in the initial states
using H-PILoT (with external prover Z3), and prove that this is the
case, then we prove that $\Phi_1$ is also invariant under updates. 
We could therefore strengthen $\Phi$ to obtain an inductive invariant
after one iteration.

\section{Systems modeled using  hybrid automata} 
\label{sect-lha}
Hybrid automata were introduced in \cite{Henzinger} to describe
systems with discrete control (represented by a finite set of 
control modes); in every control mode certain variables can
evolve continuously in time according to precisely specified rules.

\begin{definition}[\cite{Henzinger}]
A hybrid automaton 
$S = (X, Q, {\sf flow}, {\sf Inv}, {\sf Init}, E, {\sf guard}, {\sf jump})$
is a tuple consisting of: 
\begin{itemize}
\vspace{-2mm}
\item[(1)] 
A finite set $X = \{ x_1, \dots, x_n \}$ of real valued variables
(regarded as functions
$x_i : {\mathbb R} \rightarrow {\mathbb R}$) and 
a finite set $Q$ of control modes; 
\item[(2)] A family $\{ {\sf flow}_q \mid q \in Q \}$ of 
predicates over the variables in
$X \cup {\dot X}$ (${\dot X} =\{ {\dot x_1}, \dots, {\dot x_n} \}$, 
where ${\dot x_i}$ is the derivative of $x_i$)  
specifying the 
continuous dynamics in each control mode\footnote{We assume that the functions
$x_i : {\mathbb R} \rightarrow {\mathbb R}$ are differentiable during flows.}; 
a family $\{ {\sf Inv}_q \mid q \in Q \}$ of 
predicates over the variables in $X$ defining the  
invariant conditions for each control mode; 
and a family $\{ {\sf Init}_q \mid q \in Q \}$ of 
predicates over the variables in $X$, defining the initial 
states for each control mode. 
\item[(3)] A finite multiset $E$ 
with elements in $Q {\times} Q$ (the control switches), where 
every $(q, q') \in E$ is a 
directed edge between $q$ (source mode) and $q'$ (target mode); 
a family of guards 
$\{ {\sf guard}_e \mid e \in E \}$ (predicates over $X$); and a family of 
jump conditions $\{ {\sf jump}_e \mid e \in E \}$ 
(predicates over $X \cup X'$, 
where $X' =\{ x'_1, \dots, x'_n \}$ is a copy of $X$ consisting of ``primed'' 
variables). 
\end{itemize}
\end{definition}
A {\em state} of $S$ is a pair $(q, a)$ consisting
of a control mode $q \in Q$ and a vector $a = (a_1, \dots, a_n)$
that represents a value $a_i \in {\mathbb R}$ for each variable $x_i \in X$.
A state $(q, a)$ is {\em admissible} if ${\sf Inv}_q$ 
is true when each $x_i$ is replaced by $a_i$.
There are two types of {\em state change}: 
(i) A {\em jump} is an instantaneous transition 
that changes the control location and the values of variables in $X$ according
to the jump conditions;   
(ii) In a {\em flow}, the state can change due to the evolution in a 
given control mode over an interval of time: the values of the
variables in $X$  
change continuously according to the flow rules of the current
control location; all intermediate states are admissible.
A {\em run} of $S$ is a finite
sequence $s_0 s_1 \dots s_k$ of  admissible states such that
(i) the first state $s_0$ is an initial state of $S$ (the values of 
the variables satisfy ${\sf Init}_q$ for some $q \in Q$),
(ii) each pair $(s_j, s_{j+1})$ is either a jump of $S$ or the endpoints of a 
flow of $S$.

\noindent {\em Notation.} In what follows we use the following notation.
If $x_1, \dots, x_n \in X$ we denote 
the sequence $x_1, \dots, x_n$ with ${\overline x}$, the sequence 
$\dot{x}_1, \dots, \dot{x}_n$ with $\overline{\dot{x}}$, and  
the sequence of values $x_1(t), \dots, x_n(t)$ of these variables 
at a time $t$ with ${\overline x}(t)$.

\medskip
\noindent 
We identify the following verification problems: 

\medskip
\noindent {\bf Invariant checking} is 
the problem of checking whether a quantifier-free formula $\Phi$ 
in real arithmetic 
over the variables $X$ is an inductive invariant in a 
hybrid automaton $S$, i.e.: 
\begin{enumerate}
\vspace{-2mm}
\item[(1)] $\Phi$ holds in the initial states of mode $q$ for all $q \in Q$; 
\item[(2)] $\Phi$ is invariant under jumps and flows:
\begin{itemize}
\item For every flow in a mode $q$, 
the continuous variables satisfy 
$\Phi$ both during and at the end of the flow. 
\item  For every jump, 
 if the values of the continuous variables satisfy 
$\Phi$ before the jump, they satisfy $\Phi$ after the jump.
\vspace{-2mm}
\end{itemize}
\end{enumerate}

\noindent {\bf Bounded  model checking} is  
the problem of checking whether the truth of a formula 
$\Phi$ is preserved under runs 
of length bounded by $k$, i.e.: 
\begin{itemize}
\vspace{-2mm}
\item[(1)] $\Phi$ holds in the initial states of mode $q$ for every $q \in Q$; 
\item[(2)] $\Phi$ is preserved under runs of length $j$ for all $1 {\leq} j {\leq} k$.
\end{itemize}

\smallskip
\noindent A hybrid automaton $S$ is
a linear hybrid automaton (LHA) if it satisfies the following
two requirements: 
\begin{description}
\vspace{-1mm}
\item[1. Linearity] For every control mode $q \in Q$, 
the flow condition ${\sf flow}_q$, the invariant condition ${\sf Inv}_q$, 
and the initial condition ${\sf Init}_q$ are convex linear predicates.
For every control switch $e = (q,q') \in E$, the jump condition
${\sf jump}_e$ and the guard ${\sf guard}_e$ are convex linear 
predicates.
In addition, we assume that the 
flow conditions ${\sf flow}_q$ are conjunctions of {\em non-strict} 
inequalities. 
\item[2. Flow independence] For every control mode $q \in Q$, 
the flow condition ${\sf flow}_q$ is a predicate over the variables
in ${\dot X}$ only (and does not contain any variables from X).
This requirement ensures that the possible flows
are independent from the values of the variables, and depend
only on the control mode.
\vspace{-1mm}
\end{description}

\subsection{Verification, constraint generation and invariant strengthening}

We now study possibilities of verification, constraint generation and
invariant strengthening for linear hybrid automata; we assume that the
properties $\Phi$ and $\phi_{\sf safe}$ to be checked 
are convex linear predicates over $X$.

\begin{thm}[\cite{damm-ihlemann-sofronie-2011}]
The following are equivalent for any LHA: 
\begin{itemize}
\vspace{-1mm}
\item[(1)] $\Phi$ is an inductive invariant of the hybrid automaton;  

\item[(2)] 
For every $q \in Q$ and $e = (q,q') \in E$, the following formulae are 
unsatisfiable:  
\end{itemize}

\noindent $\begin{array}{@{}ll@{}}
I_q  & {\sf Init}_q \wedge \neg \Phi({\overline x}) \\[1ex] 

F_{\sf flow}(q) & \Phi({\overline x}(t_0)) \wedge {\sf Inv}_q({\overline x}(t_0)) \wedge 
{\underline {\sf flow}}_q(t_0, t) \wedge 
 {\sf Inv}_q({\overline x}(t)) \wedge 
\neg \Phi({\overline x}(t)) \wedge t \geq t_0 \\[1ex]

F_{\sf jump}(e)~~~~~ & 
\Phi({\overline x}(t)) \wedge {\sf Jump}_e({\overline x}(t), {\overline x}'(0)) \wedge  
{\sf Inv}_{q'}({\overline x}'(0)) \wedge 
\neg \Phi({\overline x}'(0))
\end{array}$

\smallskip
\noindent where if ${\sf flow}_q  =  
\bigwedge_{j = 1}^{n_q} (\sum_{i = 1}^n c^q_{ij} {\dot x}_i \leq_j
c_j^q)$ then: 

\smallskip
\noindent ${\underline {\sf flow}}_q(t, t') = 
\bigwedge_{j = 1}^{n_q}  
(\sum_{i = 1}^n c^q_{ij} (x_i' - x_i) \leq_j c_j^q (t' - t))$, where
$x'_i = x_i(t'), x_i = x_i(t)$.
\vspace{-2mm}
\label{transl-invar-par}
\end{thm}
Theorem~\ref{transl-invar-par} shows that linear hybrid automata can
be modeled as a type of constraint transition systems, in
which the updates are due to flows and jumps, and  
can be described by the formulae $F_{\sf flow}(q)$ and $F_{\sf
  jump}(e)$ above -- in $F_{\sf flow}(q)$ the value of the variable $x_i$
before the update is $x_i(t_0)$ and the value of the variable $x_i$ after 
the update is $x_i(t)$, for $t \geq t_0$.
Therefore the results in Theorems~\ref{thm-dec-inv-checking-synthesis} and~\ref{inv-gen-correctness} apply here in a simplified
form, since only the variables $X$ are updated; there are no updates 
of function symbols, so we can use Algorithm~\ref{algorithm-symb-elim} and
SEH-PILoT to check whether a formula is an inductive
invariant and, if not, for computing constraints on the parameters
under which this is guaranteed to be the case.

\subsection{Chatter-freedom and time-bounded reachability}

\noindent 
In \cite{damm-ihlemann-sofronie-2011} we also considered properties stating that for every run $\sigma$ in the automaton $S$, if 
$\phi_{\sf entry}$ holds at the beginning of the run, then 
$\phi_{\sf safe}$ becomes true in run $\sigma$ at latest at time $t$, 
i.e.\ properties of the form $\phi = \Box(\phi_{\sf entry} \rightarrow
\Diamond_{\leq t} \phi_{\sf safe})$.

\noindent For {\em chatter-free hybrid 
automata}, i.e., automata in which mode entry conditions are chosen with 
sufficient safety margin (by specifying inner envelopes described by
formulae ${\sf InEnv}_q, q \in Q$) such that a minimal dwelling time 
$\varepsilon_t$ in each 
mode is guaranteed, checking such properties 
can be reduced to bounded model checking.

\begin{defi}[\cite{damm-ihlemann-sofronie-2011}]
\label{alternative-charact}
\label{cf}
A hybrid automaton $S$ is {\em chatter-free with minimal dwelling time $\varepsilon_t$}
(where $\varepsilon_t > 0$) if:    
\begin{itemize}
\vspace{-2mm}
\item[(i)] all transitions lead to an inner envelope, i.e.\ for all $q \in Q, 
(q,q') \in E$ the following formula is valid: 

$~~~~~~~~~\forall {\overline x} ({\sf Inv}_q({\overline x}) \wedge 
{\sf guard}_{(q,q')}({\overline x}) \wedge {\sf jump}_{(q,q')}({\overline x}, {\overline x}') {\rightarrow} {\sf InEnv}_{q'}({\overline x}'));$ 
\item[(ii)] for any flow starting in the inner envelope of a mode $q$, 
 no guard of a mode switch $(q,q')$ will 
   become true in a time interval smaller than $\varepsilon_t$. 
\end{itemize}
A hybrid automaton $S$ is {\em chatter-free} iff 
$S$ is chatter-free with minimal dwelling time 
$\varepsilon_t$ for some $\varepsilon_t > 0$.  
\end{defi}
Conditions similar to chatter-freedom (e.g. finite or bounded
variability in real-time logics) were also studied e.g.\ in
\cite{Wilke-FTRTFT-1994, Maler-Nickovic-FTRTFT-2004}. 
\begin{thm}[\cite{damm-ihlemann-sofronie-2011}]
\label{entry-inner12}
Let $S$ be an LHA. \\
Condition (i) in Definition~\ref{alternative-charact} holds iff the following formula is unsatisfiable:

\smallskip
$~~~~~~{\sf Inv}_q({\overline x}) \wedge 
{\sf guard}_{(q,q')}({\overline x}) \wedge {\sf
  jump}_{(q,q')}({\overline x}, {\overline x}') {\wedge} \neg {\sf InEnv}_{q'}({\overline x}').$ 

\smallskip
\noindent Condition (ii) in Definition~\ref{alternative-charact} holds iff 
$\bigwedge_{(q,q') \in E} F_{q,q'}$ is unsatisfiable, where: 
 
\smallskip 
\noindent 
$F_{q,q'}:~~ {\sf InEnv}(x_1(0), \dots, x_n(0)) \wedge {\sf Inv}_q({\overline x}(0)) \wedge {\underline {\sf flow}}_q(0, t) \wedge$ \\
$~~~~~~~~~~~{\sf guard}_{(q, q')}(x_1(t), \dots, x_n(t)) \wedge t \leq \varepsilon_t.$

\smallskip
\noindent For any LHA $S$, checking whether $S$ is chatter-free with
  time-dwelling $\varepsilon_t$ is decidable. 
\end{thm} 
If some of the constants used in specifying these conditions are
considered to be parametric, we can use Algorithm~1 to obtain the
weakest condition on the parameters under which conditions (i)
resp.\ (ii) hold.

\subsection{Examples}
\label{example}
We consider the following (very simplified) chemical 
plant example (considered also in \cite{damm-ihlemann-sofronie-2011}), 
modeling the situation in which we control 
the reaction of two substances, and the separation of 
the substance produced by the reaction. 
Let $x_1, x_2$ and $x_3$ be variables which describe 
the evolution of the volume of substances 1 and 2, 
and the substance 3 generated from their reaction, respectively. 
The plant is described by a hybrid automaton with four modes: 

\medskip
\noindent {\bf Mode 1: Fill} In this mode the temperature is low, and hence 
the substances 1 and 2 do not react. The substances 
1 and 2 (possibly mixed with a very small quantity of substance 3) 
are filled in the tank in equal quantities up to a given error margin. This is described by the following invariants and flow conditions:

\medskip
{
\noindent \begin{tabular}{@{}ll}
${\sf Inv}_1:$ & $x_1 + x_2 + x_3 \leq L_f ~\wedge~ \bigwedge_{i=1}^3
x_i \geq 0 ~\wedge~ -\varepsilon_a \leq x_1 - x_2 \leq \varepsilon_a ~\wedge~ x_3 \leq {\sf min}$ \\
${\sf flow}_1:$ & ${\sf dmin} \leq \dot{x_1} \leq {\sf dmax} \wedge
{\sf dmin} \leq \dot{x_2} \leq {\sf dmax} \wedge \dot{x_3} {=} 0 \wedge -\delta_a {\leq} \dot{x_1} {-} \dot{x_2} {\leq} \delta_a$ \\
\end{tabular}
}

\medskip
\noindent 
If the proportion is not kept the system jumps into mode 4 ({\bf Dump});  
if the total quantity of substances exceeds level $L_f$ 
the system jumps into mode 2 ({\bf React}). 

\medskip
\noindent {\bf Mode 2: React} 
In this mode the temperature is high, and 
the substances 1 and 2 react. The reaction consumes equal quantities 
of substances $1$ and $2$ and produces substance $3$. 

\medskip 
{
\noindent \begin{tabular}{@{}ll}
${\sf Inv}_2:$ & $L_f \leq x_1 {+} x_2 {+} x_3 \leq L_{\sf overflow}
\wedge \bigwedge_{i=1}^3 x_i {\geq} 0 \wedge {-}\varepsilon_a \leq x_1 {-} x_2 \leq \varepsilon_a \wedge x_3 \leq {\sf max}$ \\
${\sf flow}_2:$ & $\dot{x_1} \leq - {\sf dmin} \wedge \dot{x_2} \leq -
{\sf dmin} \wedge \dot{x_3} \geq {\sf dmin} \wedge \dot{x_1} = \dot{x_2} \wedge \dot{x_3} + \dot{x_1} + \dot{x_2} = 0$ \\
\end{tabular}
}

\medskip
\noindent 
If the proportion between substances 1 and 2 is not kept the system jumps 
into mode 4 ({\bf Dump}); 
if the total quantity of substances 1 and 2 is below 
some minimal level ${\sf min}$ the system jumps into mode 3 ({\bf Filter}). 

\medskip
\noindent {\bf Mode 3: Filter} 
In this mode the temperature is low again and the 
substance 3 is filtered out.

\medskip
{\small 
\noindent \begin{tabular}{@{}ll}
${\sf Inv}_3:$ & $x_1 + x_2 + x_3 \leq L_{\sf overflow} ~\wedge~
\bigwedge_{i=1}^3 x_i \geq 0 \wedge -\varepsilon_a \leq x_1 - x_2 \leq \varepsilon_a ~\wedge~ x_3 \geq {\sf min}$ 
\\
${\sf flow}_3:$ & $\dot{x_1} = 0 \wedge \dot{x_2} = 0 \wedge \dot{x_3} \leq - {\sf dmin}$ \\
\end{tabular}
}

\medskip
\noindent 
If the proportion between substances 1 and 2 is not kept the system jumps 
into mode 4 ({\bf Dump}). Otherwise, if the concentration of substance 
3 is below some minimal level ${\sf min}$ the system jumps into mode 1 ({\bf Fill}). 

\medskip
\noindent {\bf Mode 4: Dump}  
In this mode the content of the tank is emptied. We 
assume that this 
happens instantaneously, i.e.\: 
${\sf Inv}_4: \bigwedge_{i=1}^3 x_i = 0$ and ${\sf flow}_4: \bigwedge_{i=1}^3 \dot{x_i} = 0$.  

\medskip
\noindent {\bf Jumps.} The automaton has the following jumps:
\begin{itemize}
\item $e_{12} = (1,2)$ with ${\sf guard}_{e_{12}} = x_1 {+} x_2 {+}
  x_3 {\geq} L_f$; 
${\sf jump}_{e_{12}} = \bigwedge_{i = 1}^3 x'_i = x_i$; 
\item $e_{23} = (2,3)$ with 
${\sf guard}_{e_{23}} = x_1 + x_2 {\leq} {\sf min}$;  
${\sf jump}_{e_{23}} =  \bigwedge_{i = 1}^3 x'_i = x_i$; 
\item $e_{31} = (3,1)$ with 
${\sf guard}_{e_{31}} {=} {-}\varepsilon_a {\leq} x_1 {-} x_2 {\leq} \varepsilon_a
\wedge 0 {\leq} x_3 {\leq} {\sf min}$; ${\sf jump}_{e_{31}} {=}  \bigwedge_{i = 1}^3 x'_i {=} x_i$; 
\item Two edges $e^1_{14}, e^2_{14}$ from $1$ to $4$,  and 
 two edges $e^1_{24}, e^2_{24}$ from $2$ to $4$,  
with:\\
 ${\sf guard}_{e^1_{j4}} = x_1 {-} x_2 {\geq} \varepsilon_a$, ${\sf guard}_{e^2_{j4}} = x_1 {-} x_2 {\leq} - \varepsilon_a$; 
 ${\sf jump}_{e^i_{j4}} = \bigwedge_{i = 1}^3 x'_i {=} 0$; 
\item Two edges $e^1_{34}, e^2_{34}$ from $3$ to $4$, with 
${\sf guard}_{e^1_{34}}$ $=$ $x_3 \leq {\sf min} \wedge x_1 - x_2 \geq \varepsilon_a$;  
${\sf guard}_{e^2_{34}} = x_3 \leq {\sf min} \wedge x_1 - x_2 \leq -
\varepsilon_a$, and ${\sf jump}_{e^i_{34}} = \bigwedge_{i = 1}^3 x'_i = 0$ for $j = 1,2$.
\end{itemize}

\

\noindent {\bf Tests with SEH-PILoT.}  We illustrate how SEH-PILoT
can be used for testing whether formulae are invariant
resp.\ for constraint generation and invariant strengthening for variants
of the linear hybrid system described before. 
We assume that ${\sf min}, {\sf max}, {\sf dmin}, {\sf dmax}, L_{\sf safe},$ $L_{f},$ $L_{\sf overflow},$ $\varepsilon_{\sf
  safe},$ $\varepsilon_{a}, \delta_a$ are parameters. 
In order to have only linear constraints, we here often assume that
${\sf dmin}, {\sf dmax}$
and ${\sf \delta_a}$ are constant (for experiments we choose ${\sf dmin} =
{\sf \delta_a} = 1, {\sf dmax} = 2$). 

\

\noindent {\bf  1. Invariant checking and constraint generation.} 
Consider
the property $\Phi = (x_1 + x_2 + x_3 \leq L_{\sf safe})$.
Without knowing the link between $L_{\sf safe}$, $L_f$ and $L_{\sf
  overflow}$ we cannot prove that $\Phi$ is an inductive invariant. 
Due to the form of jump conditions, it is easy to see that $\Phi$ is
always invariant under the jumps  $(1,2), (2, 3)$ and $(3, 1)$ which do not
change the values of the variables and it is invariant under the jumps
to mode 4 (which reset all variables to 0) iff $L_{\sf safe} \geq 0$.

\smallskip 
\noindent We used SEH-PILoT to generate the weakest constraint on 
the parameters under which $\Phi$ is invariant under flows. This
constraint is obtained by eliminating $\{ x_1, x_2, x_3, x_1', x_2',
x_3', t \}$ from the formulae ${\underline {\sf flow}}_q(0,t)$, $q \in \{ 1, 2, 3, 4 \}$.
 
\medskip
\noindent Here are the results for the flow in mode 1. 

{\scriptsize 
\begin{lstlisting}[frame=single]
tasks:
    mode1_invariant1:
        mode: GENERATE_CONSTRAINTS
        options:
            eliminate: [x1, x2, x3, x1p, x2p, x3p, t]
            slfq_query: true
        specification_type: HPILOT
        specification_theory: REAL_CLOSED_FIELDS
        specification:
            file: |
                Base_functions := {(+,2), (-,2), (*,2)}
                Extension_functions :=  {}
                Relations := {(<=,2), (<,2), (>=,2), (>,2)}
                           
                Query := % --Parameters--
                         ea > 0; 
                         % --Inv1(0)--
                         (x1 + x2) + x3 <= lf;
                         x1 >= _0; x2 >= _0; x3 >= _0;
                         x2 - x1 <= ea; x1 - x2 <= ea; x3 <= min;
                         % --flow1(0,t)--
                         x1p - x1 >= t; x2p - x2 >= t; x3p - x3 = _0;
                         (x2p - x2) - (x1p - x1) <= t;
                         (x1p - x1) - (x2p - x2) <= t;  t >= _0;
                         % --Psi(0)--
                         (x1 + x2) + x3 <= lsafe;
                         % --Inv1(t)--
                         (x1p + x2p) + x3p <= lf;
                         x1p >= _0; x2p >= _0; x3p >= _0;
                         x2p - x1p <= ea; x1p - x2p <= ea; x3p <= min;
                         % --not Psi(t)--
                         (x1p + x2p) + x3p > lsafe;
\end{lstlisting}
} 

\noindent 
SEH-PILOT gives the following output:

{\scriptsize \begin{lstlisting}[frame=single]
Metadata:
    Date: '2023-07-26 11:25:47'
    Number of Tasks: 1
    Runtime Sum: 0.4641
mode1_invariant1:
    Runtime: 0.4641
    Result: OR(min < _0, lsafe < _0, lf - lsafe <= _0, ea <= _0)
\end{lstlisting}
}

\smallskip
\noindent Thus, SEH-PILoT generates the following constraint for mode 1: 

\smallskip
$\Gamma = ({\sf min} < 0 \vee L_{\sf safe} <  0 \vee  L_f - L_{\sf
  safe} \leq  0 \vee \varepsilon_a \leq 0)$

\smallskip
\noindent If we know that ${\sf min} \geq 0, L_{\sf safe} \geq 0$ and
$\varepsilon_a > 0$ we obtain the condition $L_f \leq L_{\sf safe}$.

\

\noindent {\bf 2. Invariant strengthening.} Consider now a variant of the
hybrid system described before in which the condition $\bigwedge_{i =
  1}^3 x_i \geq 0$ was left out from the mode invariants for modes 1
and 2. We here assume in addition -- for simplifying the presentation --
that $L_f = L_{\sf overflow}$. Consider the property $\Phi = x_1 + x_2 \leq L_f$. 

\smallskip
\noindent 
We want to show that $\Phi$ holds on all runs in the automaton which
start in mode 1 in the state ${\sf Init} := x_1 {=} 0 \wedge x_2 {=} 0 \wedge x_3 {=} 0$. 
$\Phi$ clearly holds in ${\sf Init}$ if $L_f > 0$. 
However, without the additional conditions $\bigwedge_{i =
  1}^3 x_i \geq 0$ in the invariants of modes 1 and 2, we cannot prove that $\Phi$
is invariant under flows.
We try to strengthen $\Phi$ in order to obtain an inductive invariant
-- e.g.\ by obtaining an additional condition on $x_3$, 
by eliminating $\{ x_1, x_2, x_1', x_2', x_3', t \}$ from the formulae 
${\underline {\sf flow}}_q(0,t)$, $q \in \{ 1, 2, 3, 4 \}$.

{\scriptsize 
\begin{lstlisting}[frame=single]
tasks:
    mode1_invariant1:
        mode: GENERATE_CONSTRAINTS
        options:
            eliminate: [x1, x2, x1p, x2p, x3p, t]
            slfq_query: true
        specification_type: HPILOT
        specification_theory: REAL_CLOSED_FIELDS
        specification:
            file: |
                Base_functions := {(+,2), (-,2), (*,2)}
                Extension_functions :=  {}
                Relations := {(<=,2), (<,2), (>=,2), (>,2)}
                           
                Query := % --Parameters--
                         ea > 0; 
                         % --Inv1(0)--
                         (x1 + x2) + x3 <= lf;
                         x2 - x1 <= ea; x1 - x2 <= ea; x3 <= min;
                         % --flow1(0,t)--
                         x1p - x1 >= t; x2p - x2 >= t; x3p - x3 = _0;
                         (x2p - x2) - (x1p - x1) <= t;
                         (x1p - x1) - (x2p - x2) <= t;
                         t >= _0;
                         % --Psi(0)--
                         (x1 + x2)  <= lf;
                         % --Inv1(t)--
                         (x1p + x2p) + x3p <= lf;
                         x2p - x1p <= ea; x1p - x2p <= ea;
                         x3p <= min;
                         % --not Psi(t)--
                         (x1p + x2p) > lf;
\end{lstlisting}
}

\noindent 
SEH-PILOT gives the following output:

{\scriptsize \begin{lstlisting}[frame=single]
Metadata:
    Date: '2023-07-26 11:22:11'
    Number of Tasks: 1
    Runtime Sum: 0.441
mode1_invariant1:
    Runtime: 0.441
    Result: OR(x3 >= _0, min - x3 < _0, ea <= _0)
\end{lstlisting}}

\noindent Thus, SEH-PILoT generates the following constraint for mode 1:

\smallskip
$\Gamma := x_3 \geq 0 \vee x_3 > {\sf min} \vee \varepsilon_a \leq 0$.

\smallskip
\noindent Since we assume that $\varepsilon_a > 0$ and the invariant of
mode 1 contains the condition $x_3 \leq {\sf min}$, it follows that we
can consider the candidate invariant: 

\smallskip
$\Phi_1 := (x_1 + x_2 \leq L_f) \wedge x_3 \geq 0.$

\smallskip
\noindent It can be checked that $\Phi_1$ holds in the initial state
described by the formula ${\sf Init}$ and it is preserved under jumps
(under the assumption that $L_f \geq 0$) and  flows.

\medskip
\noindent {\bf 3. Constraint generation for guaranteeing chatter-freedom.}
We now analyze the property of chatter-freedom. 
Let $\varepsilon > 0$ and consider the following 
parametrically described inner envelope for mode 1: 

\medskip
{ 
${\sf InEnv}_1 := x_1 + x_2 + x_3 \leq L_{\sf safe} \wedge \bigwedge_{i = 1}^3 x_i \geq
0 \wedge |x_1 - x_2| \leq \varepsilon_{\sf safe} \wedge x_3 \leq {\sf min}$
}

\medskip 
\noindent where $0 < L_{\sf safe} \leq L_f$, $0 < \varepsilon_{\sf safe} \leq
\varepsilon_a,  {\sf dmin} < {\sf 
  dmax}$, and ${\sf min}, {\sf dmax}, {\sf dmin} > 0$. 

\medskip
\noindent We show how SEH-PILoT can be used to generate 
constraints on the parameters $L_f, L_{\sf overflow}, \varepsilon_a,
\delta_a, {\sf min}, {\sf dmin}$ and $L_{\sf safe}, \varepsilon_{\sf
  safe}$ under which a minimal dwelling time $\varepsilon$ in mode 1 is 
guaranteed. 

\medskip
\noindent We analyze the jumps $e \in \{ e_{12}, e^1_{14}, e^2_{14}\}$ from mode 1 to modes 2 and 4, 
and for each such $e$ use SEH-PILoT to generate constraints $\Gamma_e$
on the parameters under
which the following formula is unsatisfiable:

\medskip
${\sf InEnv}_1({\overline x}) \wedge {\sf Inv}_1({\overline x}) \wedge
{\underline {\sf flow}}({\overline x}, {\overline x}', t) \wedge {\sf
  guard}_{e}({\overline x}') \wedge t < \varepsilon.$

\medskip
\noindent 
Below is the test for the jump from mode 1 to mode 4 with guard $x_1 -
x_2 \geq \varepsilon_a$.
We prepared two input files: in the first one we generate constraints
without simplifying them; the second one with simplification.

\medskip
{\scriptsize 
\begin{lstlisting}[frame=single] 
tasks:
    example_Damm_1:
        mode: GENERATE_CONSTRAINTS
        options:
            eliminate: [x1, x2, x3, x1p, x2p, x3p, t]
            slfq_query: true
        specification_type: HPILOT
        specification_theory: REAL_CLOSED_FIELDS
        specification: &spec_Damm_2
            file: |
                Base_functions := {(+,2), (-,2), (*,2)}
                Extension_functions :=  {}
                Relations := {(<=,2), (<,2), (>=,2), (>,2)}
                           
                Query := % --Inner envelope-- 
                         (x1 + x2) + x3 <= lsafe;
                         _0 <= x1;  _0 <= x2;  _0 <= x3; x3 <= min;
                         -esafe <= x1 - x2;  x1 - x2 <= esafe;
                         % --Parameters--
                         _0 < lsafe; lsafe < lf; _0 < esafe; esafe < ea;
                         % --flow1(0,t)--
                         _0 < t;  x3p = x3;
                         dmin * t <= x1p - x1; x1p - x1 <= dmax * t;
                         dmin * t <= x2p - x2; x2p - x2 <= dmax * t;
                         (x1p - x2p) - (x1 - x2) <= da * t;
                         (x2p - x1p) - (x2 - x1) <= da * t;
                        % --Guard of first transition 1->4 true--
                         ea <= x1p - x2p;
                        % --Dwelling time too short--
                         t <= epsilon;
\end{lstlisting}}

\

\noindent The constraint generated without simplification is really large. 
Below is the version obtained after simplification with
SLFQ. (Note that the time needed for quantifier elimination and simplification for non-linear
formulae is quite high.)

\

{\scriptsize 
\begin{lstlisting}[frame=single]
Metadata:
    Date: '2023-07-27 15:04:16'
    Number of Tasks: 1
    Runtime Sum: 21.6991
example_Damm_1:
    Runtime: 21.6991
    Result: OR(min < _0, lsafe <= _0, lf - lsafe <= _0, esafe <= _0, 
        epsilon <= _0, ea - esafe <= _0, dmax - dmin < _0, da < _0,  
      (((dmax * epsilon) - (dmin * epsilon)) - ea) + lsafe < _0, 
      (((dmax * epsilon) - (dmin * epsilon)) - ea) + esafe < _0, 
      ((da * epsilon) - ea) + lsafe < _0, 
      ((da * epsilon) - ea) + esafe < _0)
\end{lstlisting}}

\

\noindent Under the assumption that ${\sf min} \geq 0, {\sf dmax} \geq
{\sf dmin}$, $0 < L_{\sf safe} < L_f$, $0 < \varepsilon_{\sf safe} <
\varepsilon_a$, $\delta_a > 0$ we obtain: 

\

\noindent $\Gamma_{e^1_{14}}: L_{\sf safe} < \varepsilon_a - (({\sf dmax} -
{\sf dmin}) * \varepsilon) \vee 
e_{\sf safe} < (\varepsilon_a - ({\sf  dmax} - {\sf dmin}) * \varepsilon) 
\vee$  \\
$~~~~~~~~~ L_{\sf safe} < (\varepsilon_a - \delta_a * \varepsilon)  \vee
\varepsilon_{\sf safe} < (\varepsilon_a - \delta_a * \varepsilon)$

\

\noindent Thus, SEH-PILoT generates the following
constraints (simplification with SLFQ is crucial
as before simplification the formulae are very long); we left out
the parts of the conjunction which are negations of conditions on the parameters.

\

{
\noindent $\begin{array}{ll}
\Gamma_{e^1_{14}}, \Gamma_{e^2_{14}}: & L_{\sf safe} < \varepsilon_a - (({\sf dmax} -
{\sf dmin}) * \varepsilon) \vee 
e_{\sf safe} < (\varepsilon_a - ({\sf  dmax} - {\sf dmin}) * \varepsilon) 
\vee  \\
& 
L_{\sf safe} < (\varepsilon_a - \delta_a * \varepsilon)  \vee \varepsilon_{\sf safe} < (\varepsilon_a - \delta_a * \varepsilon)   \\


\Gamma_{e_{12}}: & L_{\sf safe} < (L_f - (2 * {\sf dmax}) * \varepsilon).
\end{array}$}
 
\

\noindent These and further tests can be found under: \\
\url{https://userpages.uni-koblenz.de/~sofronie/tests-symbol-elimination/}

\section{Conclusions}
\label{conclusions}

In this paper we presented an approach to the verification of parametric
systems based on hierarchical reasoning and hierarchical symbol
elimination as well as an ongoing implementation. The work in this
area was strongly influenced by the collaborations within the AVACS
project and in particular by the collaboration with Werner Damm, to
whom we dedicate this article.

In future work we plan to extend the implementation of the
hierarchical symbol elimination with further optimizations proposed in 
\cite{peuter-sofronie-cade2019}, use these optimizations for making invariant
strenghtening more efficient, and compare the results with the
orthogonal results proposed in \cite{Shoham17}.
We would like to analyze the applicability of these
ideas to the verification of systems consisting of a parametric number
of similar, interacting components, in the context of modular 
approaches to verification -- for instance approaches we considered in
\cite{IFM-2010} and \cite{damm-horbach-sofronie-frocos2015}, 
either using locality of logical theories for establishing a small model property,
or using methods proposed in \cite{Sofronie97, Sofronie-getco}.

\smallskip
\noindent {\bf Acknowledgments:} We thank the reviewers for
their helpful comments.

{\bibliographystyle{abbrv}
\bibliography{references}
}

\end{document}